\documentclass[aps,prl,twocolumn,showpacs,superscriptaddress,groupedaddress]{revtex4}  

\def\alens{a}
\def\Tsls{T_{s}}
\usepackage{graphicx}
\usepackage{siunitx}

\newcommand{\NRAYS}{$10^8$}
\newcommand{\aj}{AJ}
\newcommand{\apjs}{ApJS}
\newcommand{\aap}{A\&A}
\newcommand{\pasj}{Publ. Astron. Soc. Jpn.}
\newcommand{\mnras}{Mon. Not. R. Astron. Soc,}
\newcommand{\physrep}{Phys. Rep.}
\newcommand{\rmxaa}{Rev. Mexicana Astron. Astrofis.}
\newcommand{\na}{New Astronomy}

\begin{document}
\title{The Distortion of the Cosmic Microwave Background by the Milky
  Way} \date{\today} 
\author{Benjamin Czaja, Benjamin C.\ Bromley
\\
University of Utah, Department of Physics and Astronomy}

\begin{abstract}
The Milky Way can act as a large-scale weak gravitational lens of the
cosmic microwave background (CMB). We study this effect using a photon
ray-tracing code and a Galactic mass distribution with disk, bulge and
halo components. For an observer at the Sun's coordinates in the
Galaxy, the bending of CMB photon paths is limited to less than one
arcsecond, and only for rays that pass within a few degrees of the
Galactic Center. However, the entire sky is affected, resulting in
global distortions of the CMB on large angular scales. These
distortions can cause the low-order multipoles of a spherical harmonic
expansion of the CMB sky temperature to leak into higher-order
modes. Thus the component of the CMB dipole that results from the
Local Group's motion relative to the local cosmic frame of rest
contributes to higher-order moments for an observer in the solar
system.  With our ray-tracing code we show that the phenomenon is not
sensitive to the specific choice of Galactic potential. We also
quantitatively rule it out as a contributor to CMB anomalies
such as power asymmetry or correlated alignment of low-order multipole
moments.

\end{abstract}
\maketitle

\section{Introduction}

As an echo of the hot big bang, fluctuations in the cosmic microwave
background are snapshots of seeds of cosmic structure, including
galaxies, galaxy clusters and the large-scale network of voids,
filaments, and superclusters. NASA's {\em Cosmic Background Explorer}
first detected these fluctuations, which appear as angular
anisotropies on the plane of the sky \cite{cobe92}. More recently the
{\em Wilkinson Microwave Anisotropy Probe } \cite{wmap13} and the {\em
  Planck Collaboration} \cite{planck13} have mapped out these
temperature fluctuations in exquisite detail, allowing for precise
assessment of cosmological parameters such as the global mass density
and the baryon fraction \cite{teg04}.

Distortions of the cosmic microwave background by sources between the
surface of last scattering and an observer in the present day can
measurably alter the primordial CMB. For example, the Sachs-Wolf
effect of gravitational red shifting that results from evolving
large-scale structure can generate anisotropies in the CMB sky
temperature \citep{sw67}. Additional anisotropies arise from the
Sunyaev-Zeldovich effect of photon scattering by hot gas in galaxy
clusters \citep{sz72}. Weak gravitational lensing by smaller-scale
structures, including individual distant galaxies, can produce
fluctuations as well. \citep[for a review, see][]{lew06}.

The CMB also exhibits temperature anisotropies on the largest
scales. In addition to the Doppler affect that produces the dominant
dipole feature \citep{lin96}, the CMB has quadrupole and octupole
modes that may be aligned as compared with a statistically
isotropic field \citep{teg03}.  Furthermore, the statistics of
temperature fluctuations within separate hemispheres of the sky are
significantly different, yielding asymmetric power spectra
\citep{eri04,han04,hemiplanck}. These large-scale anomalies challenge
a fundamental assumption of modern cosmology, that the universe is
statistically isotropic and homogeneous.

Here we focus on gravitational lensing of the CMB by the largest
structure in the sky, our own Galaxy.  This effect will be weak:
Indeed if the CMB were perfectly isotropic, and if the Galaxy and
observers within it were at rest in the CMB's reference frame, then
lensing would produce no temperature anisotropies. However, the Galaxy
is moving with respect to the CMB, and observers in the Galactic rest
frame see anisotropies from the relativistic Doppler shift of CMB
photons. In terms of spherical harmonic expansions of the CMB
temperature maps, the dipole is most prominent, with weak
contributions from higher-order modes \citep{kam03}. It is this
Doppler shifted map that Galactic lensing distorts. Since the dipole
is strong, the distortions might generate higher-order multipoles at
measurable levels.

In this paper, we select models of the Galaxy's mass distribution and
present a ray-tracing algorithm to track photon trajectories in these
models (\S2).  We then consider how a uniform CMB signal is affected
by the motion of the Galaxy relative the CMB's frame of reference, and
the Galactic potential (\S3).  We show that lensing mixes
the dipole signal with higher-order spherical harmonic modes of the
CMB fluctuations in the plane of the sky.  Finally we discuss our
results in the context of experimental measurements of the CMB.

\section{Weak lensing by the Galaxy}

\subsection{Ray Tracing through the Milky Way}

To map photon trajectories through a gravitational potential $\Phi$, we
start at the observer location near the Sun, taken to be 8.0 kpc 
from the Galactic Center in the plane of the disk. We ray-trace back
in time, using initial velocities aimed at points on a regular
grid of Galactic latitude and longitude coordinates in the observer's
sky. A 4-th order Richardson extrapolation integrator
\citep{bro06} solves the spatial part of the photon geodesic
equation in the limit of a weak, static source of gravity,
\begin{equation}
    \ddot{\vec{x}} = -2\vec{\nabla}_\perp\Phi ,
\end{equation} 
where the spatial derivative is the component of the gradient that is
perpendicular to the photon velocity. Integration ends when a photon
has reached a distance of several hundred kpc or more from the
Galactic Center.

We select the gravitational potential $\Phi$ from a suite of published models:
\citet[]{ken08}, \citet[]{pac90},
\citet[]{jon95}, \citet[]{dau95}, and \citet[]{all91}.  These models
all contain disk (Miyamoto-Nagai \cite{mn75}), bulge (Hernquist
\cite{hq90}, Plummer \cite{pl11}), and halo components
(Navarro-Frenk-White \cite{nfw97}). Where the mass is divergent, we
take the extent of the models to be 300~kpc. Our results are not sensitive
to this choice, or to the radial limit of photon integration. While we do
find some sensitivity to specific models, our overall conclusions hold
for all models. Unless otherwise specified, we adopt the form of
$\Phi$ from \citet[]{ken08}.

\subsection{Temperature maps and spherical harmonic coefficients}

To create a map of sky brightness, we sample the full grid of
angular coordinates, with $N_b \times N_\ell$ points in latitude $b$
and longitude $\ell$.  A ``pixel'' with index $ij$ has sky location
$(b_i,\ell_j)$ with $0\le i<N_b$ and $0\le j<N_\ell$; the unit vector
$\hat{n}_{ij}$ aimed out from the observer in that direction gives the
initial direction of the photon ray associated with that pixel.
Lensing can change the final direction of that ray, given by
$\hat{n}^\prime_{ij}$; a deflection angle
\begin{equation}
\label{eq:deflecang}
\Psi_{ij} =\arccos(\hat{n}^\prime_{ij}\cdot\hat{n}_{ij})
\end{equation}
provides a measure of the lensing effect.  Fig. \ref{fig:aitoff}
provides an illustration, showing deflection angles as a function of
sky position. In this case,
the maximum deflection angle is limited to approximately one
arcsecond.

The sky brightness measured at a pixel in direction $\hat{n}$ depends
on the brightness of the primordial CMB in direction
$\hat{n}^\prime$. Modifications to this primordial signal include a
Doppler shift from the peculiar motion of the Galaxy, a gravitational
redshift from the Galactic potential, and a cosmological redshift.  A
single factor $g$, the ratio of an observed photon's frequency $\nu$
relative to its frequency $\nu^\prime$ at the surface of last
scattering, can account for both the cosmological and gravitational
redshifts. We do not attempt to disentangle them --- the cosmological
redshift will be overwhelmingly dominant --- because they do not affect
the anisotropy of sky temperature. Since the CMB's specific intensity
$I_\nu$ corresponds to a Planck function $B_\nu(T)$ at temperature
$T$, and since $I_\nu/\nu^3$ is a Lorentz invariant (conservation of
photon number along a ray), it follows that
\begin{equation}
I_\nu(T) = B_\nu(T) = B_\nu(g \Tsls),
\end{equation}
where $\Tsls$ is the CMB temperature at the surface of last scattering.

Compared to the redshift of CMB photons, the Doppler effect of the
Galaxy plowing through the CMB is more complicated.  We view it as the
result of a boost from the frame of the CMB to the frame of the
Galaxy, in which we calculate the gravitational lensing. For
definiteness, we assume that the velocity of the Galaxy relative to
the CMB, $\vec{V}$, has a magnitude of 627~km/s and direction
$(\ell,b)=(30^\circ, 276^\circ)$, corresponding to the Local Group
\citep{kog93}.  Taking into account the lensing of the CMB, the
observed blackbody temperature in a specific direction $\hat{n}$ is
\begin{eqnarray} 
T(\hat{n}) & = & \frac{g\Tsls\sqrt{1-V^2/c^2}}
{1-\hat{n}^\prime\cdot\vec{V}/c} \\
\ & \approx & g\Tsls \left[1 
+ \frac{V}{c} \cos\theta 
+ \frac{V^2}{2c^2} \cos{2\theta}
+ \mathcal{O}\left(\frac{V^3}{c^3}\right)\right]
\label{eq:T}
\end{eqnarray}
where $\theta$ is the angle between $\hat{n}^\prime$ and $\vec{V}$.
With the ray-tracer to calculate $\hat{n}^\prime$, we find $T(\hat{n})$
for each pixel and build up a temperature map for any Galactic potential,
covering the full sky with up to \NRAYS\ pixels.

To measure the impact of lensing on the CMB, we work with 
conventional spherical harmonic expansion coefficients,
\begin{equation}\label{eq:alm}
\alens_{\ell,m} =  \int^{2\pi}_0 \int^{\pi}_0 
T(\hat{n}) Y_{\ell,m}^* (\theta,\phi) \sin(\theta) \,d\theta d\phi
\end{equation}
where the sky temperature $T$ is measured along a lensed ray.  We
estimate these coefficients---and their values in the limit of no
lensing---from the pixel maps using a third-order accurate Newton-Cotes
integration scheme. Our results below are reported in terms of
$C_\ell$, defined as the absolute square of the expansion coefficients
for a particular $\ell$, averaged over $m$ values.

\section{Results}

\begin{figure}[!h]
\centering
\includegraphics[scale=0.15]{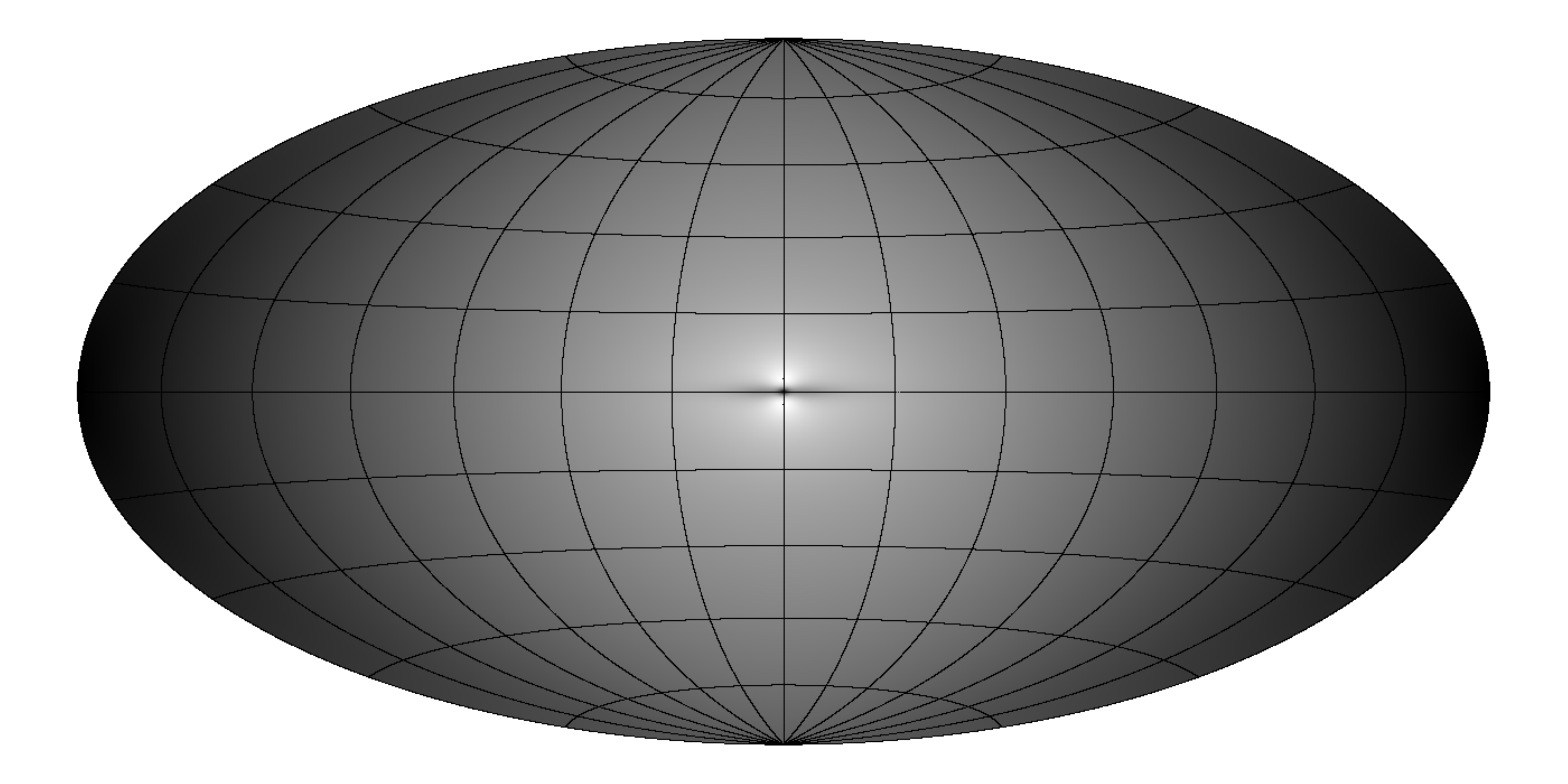}
\caption{An Aitoff projection of lensed light rays by the Milky Way
  forming a pattern across the entire sky. The area on the sky
  where the light rays were lensed the most appears lighter, where the
  light rays were lensed the least the appears darker. The maximum
 deflection angle is approximately 0.8$''$}
\label{fig:aitoff}
\end{figure}

In principle the lensing of light rays over the sky maps out the
gravitational potential of the Milky Way.  Since the sun is not
located at the center of the Galaxy we observe a distinct lensing
pattern, as in Fig. \ref{fig:aitoff}. The strongest lensing is in
the direction of the bulge of the Milky Way, a region where stars are
densely clustered.  However lensing still occurs across most of the
sky as a result of the dark matter halo. When the Galaxy is moving
with respect to the CMB rest frame, the resulting dipole becomes a
``background'' signal that can leak into other CMB modes as a result
of the combination of the sun's location away from the Galactic Center
and the small-magnitude but large-scale deflection of light rays.

\begin{figure}[!h]
\centering
\includegraphics[scale=0.40]{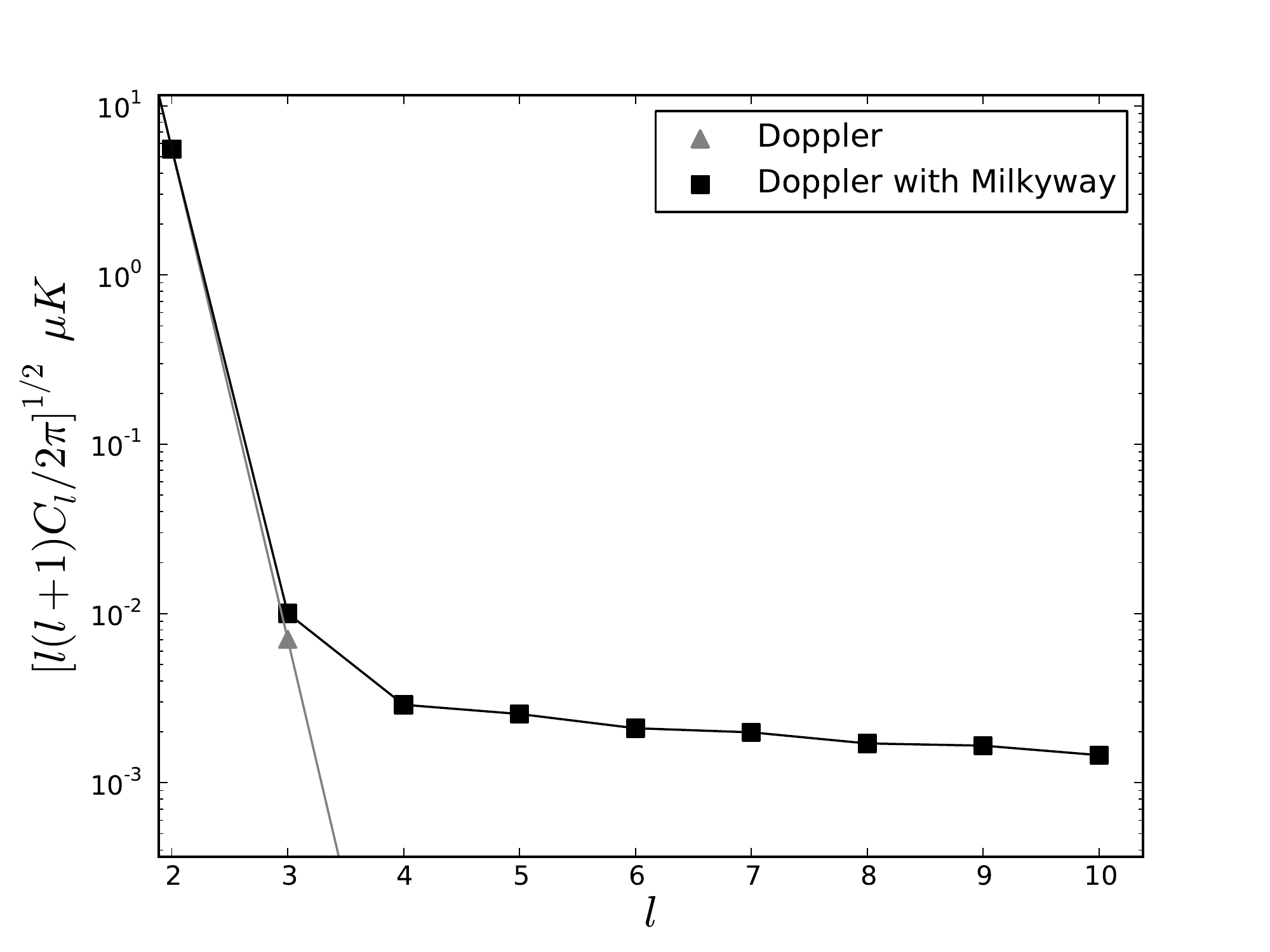}

\caption{Amplitude of spherically-averaged large-angle anisotropies
  due to the Doppler effect from the motion of our Galaxy (triangles),
  and further including the effect of gravitational lensing by the
  Galaxy (squares), as a function of angular scale (given in terms of
  multipole index $\ell$). For the octupole and higher-order
  terms, the distortions from lensing dominate the Doppler effect.}
\label{fig:result}
\end{figure}

Fig. \ref{fig:result} compares the Doppler effect both with and
without lensing. The two cases have similar mode amplitudes up to the
octupole terms, however, higher-order modes show remarkable
differences. The fall-off in the lensed CMB spectrum is slow but
steady with increasing $\ell$, while the Doppler effect on its own
plummets, as expected from Eq.~\ref{eq:T}.  Table~I gives a more
quantitative comparison of the low-$\ell$ angular power spectrum in
terms of the amplitude spectral density, $\propto \sqrt{C_\ell}$.

\begin{table}[!ht]
\caption{
The temperature anisotropies induced by the Doppler effect from the
Galaxy's motion, with and without a contribution from lensing, on
large angular scales. The first column is the multipole index, while
the angle-averaged mode amplitudes for a Doppler only model (no
lensing), and the Doppler effect along with lensing are in the middle
and right columns, respectively.  } \centering
\begin{tabular}{c c c}
\hline \hline 
mode & \multicolumn{2}{c}{amplitude (${[l(l+1)C_l/2\pi]^{1/2} K}$)} \\
${\ell}$ & \ \ \ Doppler only \ \ \  & \ \ \ Doppler with lensing \ \ \ \ \\
\hline
1 & 3.83 ${\cdot}$ 10$^{-3}$ &  3.83 ${\cdot}$ 10$^{-3}$ \\
2 & 5.55 ${\cdot}$ 10$^{-6}$ &  5.55 ${\cdot}$ 10$^{-6}$ \\
3 & 7.04 ${\cdot}$ 10$^{-9}$ &  9.52 ${\cdot}$ 10$^{-9}$ \\
4 & $<$10$^{-12}$ &  2.49 ${\cdot}$ 10$^{-9}$ \\
5 & $<$10$^{-12}$ &  2.14 ${\cdot}$ 10$^{-9}$ \\
6 & $<$10$^{-12}$  &  1.81 ${\cdot}$ 10$^{-9}$ \\
7 & $<$10$^{-12}$  &  1.67 ${\cdot}$ 10$^{-9}$ \\
8 & $<$10$^{-12}$  &  1.48 ${\cdot}$ 10$^{-9}$ \\
9 & $<$10$^{-12}$  &  1.40 ${\cdot}$ 10$^{-9}$ \\
10 & $<$10$^{-12}$  &  1.27 ${\cdot}$ 10$^{-9}$ \\
\hline 

\end{tabular}
\end{table}

\begin{figure}[!Ht]
\centering
\includegraphics[scale=0.4]{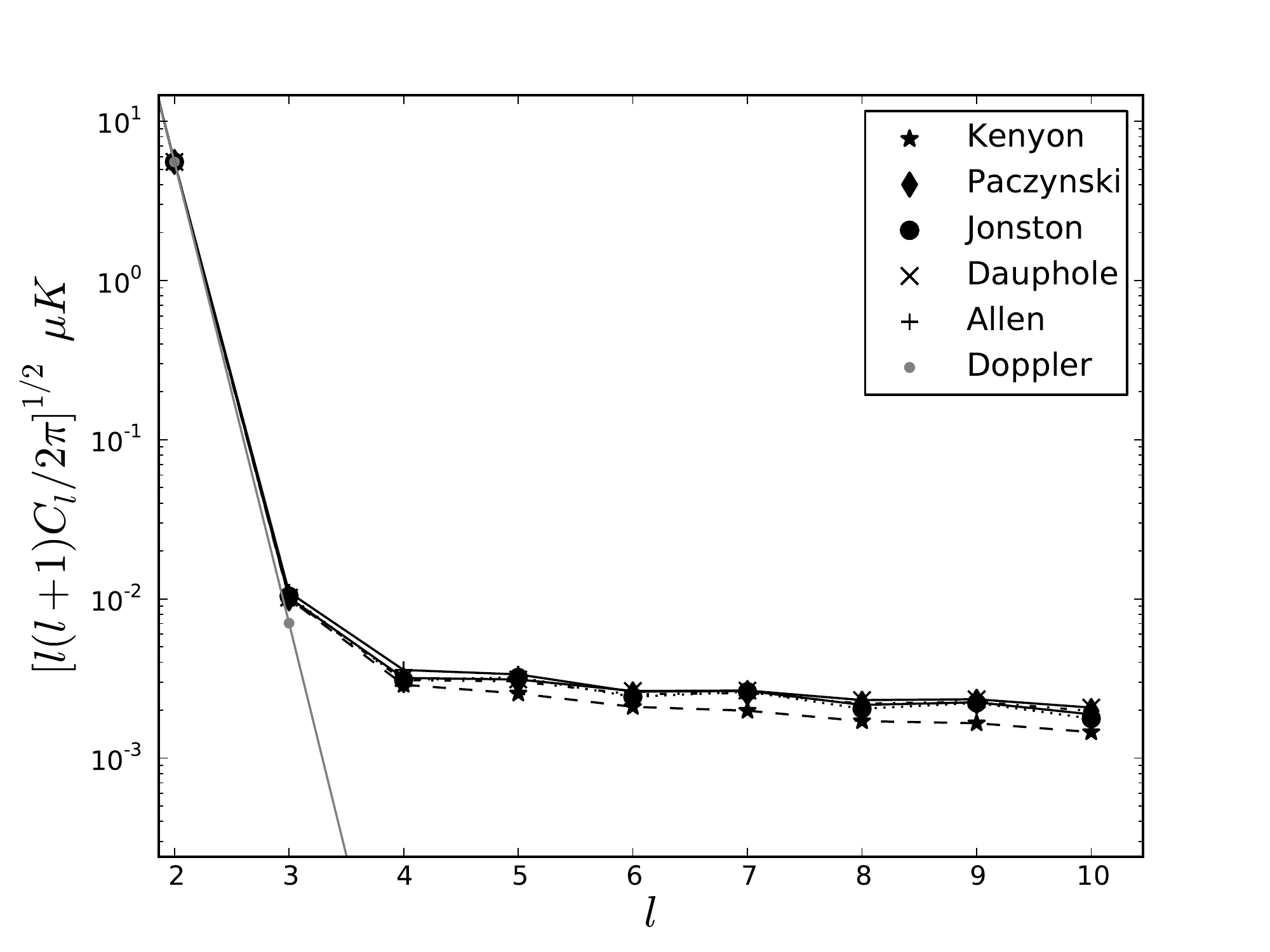}
\caption{The amplitude spectrum of fluctuations that results from the
  Doppler effect and gravitational lensing as predicted in a set of
  Galactic potential models. The models are labeled Kenyon (stars),
  Paczynski (diamonds), Jonston (black circles), Dauphole (x), and
  Allen (crosses), as described in the text.  For comparison, the
  Doppler-only model (grey circles) is also shown falling off abruptly
  with angular scale. The Galactic mass models all give results that
  are similar to one another and all generally overwhelm the Doppler-only
  spectrum at $\ell > 3$.
}
\label{fig:diff}
\end{figure}

Fig.~\ref{fig:diff} demonstrates the sensitivity of the lensing effect
to the model of the Galactic potential.  The overall power spectrum is
similar in all cases. Differences between the models, while small,
illustrate that the Milky Way lensing effect may provide some level of
discrimination between them, at least in the idealized setting
afforded by our ray-tracing experiments. Even if models of the Galaxy
are not distinguishable, broader properties of the Galaxy might be
assessed. For example the strength of the low-$\ell$ modes scale with
Galactic mass, $M$, assuming that the radial extent of the Galaxy is
fixed. Then the spectrum of mode amplitudes for $\ell \ge 4$ will
depend on mass ,scaling roughly as $10^{-8}/\ell \times
M/(10^{10}~M_\odot)$.

We also determine the effect of a sky mask that excludes regions near
the Galaxy's disk plane where gas and dust can contaminate the CMB
signal. Specifically we mask out low Galactic latitudes $|b| \leq
10^\circ$, setting the sky temperature to zero in the masked
region. In this way we get a general idea of how a Galactic cut may
affect the strength of the lensed CMB.  Table~II shows the difference
in spectral amplitude ($\propto$ the square root of the power, as in
Table~I) between the lensed and unlensed cases in the presence of the
mask. We see that the effect of the cut generates a signal at a level
of $10^{-4} \mu K$ for low $\ell>3$, an order of magnitude below the
unmasked signal. This reduction is mitigated somewhat if we estimate
the power from the cut-sky map by renormalizing to the fraction of the
sky that is retained in the map.

\begin{table}[!ht]
\caption{The contribution to the amplitude spectrum by gravitational
  lensing when a low-latitude Galactic cut is applied.  The first
  column is the multipole index, while the second column is the
  difference between the multipole amplitude with and without
  lensing. Here, all amplitudes are measured after
  we mask the low-latitude region by setting the sky temperature to zero
  wherever $|b| \leq 10^{\circ}$.
} \centering
\begin{tabular}{c c}
\hline \hline
mode & \ \ \ \ amplitude difference\ \ \ \  \\ 
 ${\ell}$ & ${[\ell(\ell+1)C_\ell/2\pi]^{1/2} K}$  \\
\hline
1 & 8.59 ${\cdot}$ 10$^{-10}$ \\
2 & 1.70 ${\cdot}$ 10$^{-10}$ \\
3 & 2.31 ${\cdot}$ 10$^{-10}$ \\
4 & 1.34 ${\cdot}$ 10$^{-10}$ \\
5 & 5.10 ${\cdot}$ 10$^{-10}$ \\
6 & 1.28 ${\cdot}$ 10$^{-10}$ \\
7 & 5.34 ${\cdot}$ 10$^{-10}$ \\
8 & 1.11 ${\cdot}$ 10$^{-10}$ \\
9 & 4.90 ${\cdot}$ 10$^{-10}$ \\
10 & 8.60 ${\cdot}$ 10$^{-11}$ \\
\hline 
\end{tabular}
\end{table}

\section{Discussion}

Here we calculate the deflection of cosmic photons by the gravity of
the Milky Way. All-sky deflection maps show how a background reference
signal might be distorted, although the angles involved are small
(less than an arcsecond).  Nonetheless, if some such background
exists, as in the case of of distant galaxies in the study of
cosmological weak lensing, then the Milky Way's full gravitational
potential could be revealed.

We illustrate this effect with the CMB temperature. The dipole signal
from the Galaxy's motion through the CMB serves as a reference signal,
and distortions from gravitational lensing are measured in terms of
leakage into quadrupole and higher-order modes. We demonstrate that
the effect is orders of magnitude too small to account for anomalies
like the hemisphere power asymmetry or quadrupole-octupole alignment.
Indeed, the cosmological signal in the CMB sky temperature overwhelms
the lensing effect described here. However, as technology improves,
power from lensing might contribute at a measurable level.

Even with existing data, we can use the lensing effect reported here
to place very crude astrophysical constraints. Galaxy potential models
in which the mass normalization is a parameter give a limit of $\sim
10^{15}$~M$_\odot$ for the total mass to the Milky Way.  While not
useful in terms of understanding our own Galaxy, this exercise
reflects the well-known idea that CMB lensing can yield mass estimates
for distant galaxy clusters \cite{sel00,vale04,yoo08}. Our ray-tracing
code suggests that the lensing effect may help place limits on nearby
objects as well. Perhaps Andromeda or the Virgo cluster might offer
opportunities closer to home.

\begin{acknowledgments}
We thank an anonymous referee for suggestions that improved the focus
and presentation of the manuscript. We are grateful to the University of
Utah for support through the Undergraduate Research Opportunities
Program and to NASA for a generous allotment of computer time on the
‘discover’ cluster.
\end{acknowledgments}
\bibliographystyle{plain}

\end{document}